\documentstyle[12pt,psfig,aaspp4]{article}
\begin{document}

\title{Using Astrometry to Deblend Microlensing Events}

\author{David M. Goldberg}
\affil{Princeton University Observatory, Princeton, NJ 08544-1001}
\affil{email: goldberg@astro.princeton.edu}

\begin{abstract}
We discuss the prospect of deblending microlensing events by observing
astrometric shifts of the lensed stars.  Since microlensing searches are
generally performed in very crowded fields, it is expected that stars will be
confusion limited rather than limited by photon statistics.  By performing
simulations of events in crowded fields, we find that if we assume a dark lens
and that the lensed star obeys a power law luminosity function, $n(L)\propto
L^{-\beta}$, over half the simulated events show a measurable astrometric
shift.  Our simulations included $20000$ stars in a $256\times 256$ Nyquist
sampled CCD frame.  For $\beta=2$, we found that $58 \%$ of the events were
significantly blended $(F_{\ast}/F_{tot}\leq 0.9)$, and of those, $73 \%$ had a
large astrometric shift $(\geq 0.5 \ {\rm pixels})$. Likewise, for $\beta=3$,
we found that $85 \%$ of the events were significantly blended, and that $85
\%$ of those had large shifts.  Moreover, the shift is weakly correlated to the
degree of blending, suggesting that it may be possible not only to detect the
existence of a blend, but also to deblend events statistically using shift
information.
\end{abstract}


\section{Introduction}

There has been significant discussion about blending in gravitational
microlensing events (Di Stefano \& Esin 1995; Alard 1996b; Wo\'{z}niak \&
Paczy\'{n}ski 1997; Han 1997b).  A blended microlensing event occurs when a
lensed object cannot be resolved from other nearby objects in the field of
view.  In general, this can be caused by contributions from a bright lensing
star, a binary companion to the lensed star, or a crowded field.  Microlensing
searches are typically conducted in very crowded fields in order to maximize
the frequency of detections, and thus we shall primarily deal with the latter
in this paper.

An observed star may be comprised of several contributing stars, and its total
brightness can be determined by fitting a point spread function (PSF) to the
light distribution on the CCD.  The individidual brightnesses of the
contributing stars, however, are unknown.  We define the blending parameter,
$f$, to be the ratio of the (unknown) brightness of the lensed star to the
total measured brightness of the observed star.  That is:
\begin{equation}
f\equiv \frac{F_{s0}}{\sum_{i}F_{i}}=\frac{F_{s0}}{F_{0}}\ ,
\label{eq:fdef}
\end{equation}
where $F_{s0}$ is the baseline flux from the lensed star, $F_{i}$ are the
fluxes from each contributing star in the observed image, and $F_{0}$ is the
measured baseline flux of the image.  Since only one star within the blend will
be lensed, if the lensed star is magnified by some amount, $A$, then the total
image will appear to be magnified by:
\begin{equation}
A_{obs}=1+f(A-1)\ .
\label{eq:magline}
\end{equation}
We draw a distinction between the actual magnification, $A$, and the observed
magnification, $A_{obs}$.  The former is not known directly, and can only be
inferred for a given value of $f$.  The latter is given by the ratio of
observed flux at some time to the baseline level.

If we make the simplifying assumption that the source and the lens are a point
source and a point mass, respectively, and that the two are moving at constant
speed relative to each other, then $A$ evolves in a very straightforward way
(Paczy\'{n}ski 1986; for comprehensive reviews, see also Paczy\'{n}ski 1996;
Alcock 1997):
\begin{eqnarray}
A(u)=\frac{u^{2}+2}{u(u^{2}+4)^{1/2}}\ ; & &  u^{2}(t)=
u_{min}^{2}+\left(\frac{t-t_{max}}{t_{0}}\right)^{2}\ ,
\end{eqnarray}
where $u$ is the angular distance of the lensed star from the lensing mass
in terms of the Einstein radius, $u_{min}$ is the impact parameter in the
same units, $t_{max}$ is the time of maximum brightness and $t_{0}$ is the
characteristic time of the event:
\begin{equation}
t_{0}=0.214\ {\rm yr} 
\left(\frac{M}{M_{\odot}} \right)^{1/2}
\left(\frac{D_{d}}{10 {\rm kpc}}  \right)^{1/2}
\left(1-\frac{D_{d}}{D_{s}} \right)^{1/2}
\left(\frac{\ 200 {\rm km\ s}^{-1}}{V} \right)
\label{eq:t0}\ .
\end{equation}
where $M$ is the mass of the lens, $D_{d}$ is the distance from the observer to
the lens, $D_{s}$ is the distance from the observer to the lensed star, and $V$
is the relative velocity of the two, projected into the plane of the lens.

Thus, a single lens microlensing event can be described by five parameters:
$F_{0}$, $t_{0}$, $t_{max}$, $u_{min}$, and $f$.  The total baseline flux,
$F_{0}$ can be well measured with many observations of the star in the unlensed
state, and symmetry shows that the time of peak amplification, $t_{max}$ is
unaffected by blending.  The three parameters, $f$, $t_{0}$, and $u_{min}$ are
not algebraically degenerate, but nevertheless form {\em almost} identical
light curves.  If the photometric measurements of the event are accurate
enough, it may be possible to determine these three parameters by performing a
best fit to the shape of the observed light curve.  In fact, this has been
successfully applied.  Alard (1996b) shows OGLE \#5 to be strongly blended and
the MACHO collaboration (Alcock et al. 1996) find 3 of the 9 observed events in
the LMC to have significant blends.  However, Wo\'{z}niak and Paczy\'{n}ski
(1997) show that for reasonable errors in the measurement of the light curve,
$f\ll 1$, $t_{0}$, and $u_{min}$ form a degenerate set of models such that:
\begin{eqnarray}
f'=fC\ ; & u_{min}'=u_{min}C\ ; & t_{0}'= t_{0}C^{-1} \ ,
\end{eqnarray}
where $C$ is an arbitrary constant.

If we falsely assume a given event with blending fraction, $f$, to be
unblended, then we may, in fact, be able to find a good fit of parameters in
the least squares sense, but actually underestimate $t_{0}$ by a factor of
$f^{-1}$.  If we have a good model of the geometry and velocity distribution of
the system, then this causes us to underestimate the mass by a factor of
$f^{-2}$ (equation~\ref{eq:t0}).  For a large blending factor, this could
result in mistaking a brown dwarf for a main sequence star.

There are some cases such as DUO \# 2 (Alard et al. 1995; Alard 1996a)
OGLE \# 7 (Udalski et al. 1994a), and MACHO LMC \# 9 (Bennett et
al. 1997) in which a binary system lenses a star (see Mao \& Di
Stefano 1995 for details).  In this case, the curves for different
values of $f$ are distinguishable.  Moreover, Witt \& Mao (1995)
show that the true magnification of a lensed source within the
caustics of a binary must be $\geq 3$, placing a strict upper limit on
$f$ for a given value of $A^{max}_{obs}<3$.  Indeed, substantial
blending was found in all three cases, and may lead one to the
suspicion that most single events toward the galactic bulge and the
LMC are significantly blended as well.

Several methods have been suggested for detecting blends in single
lens events.  Some (Alard 1996b; Buchalter et al. 1996; Han 1997b)
suggest that it may be possible to use the color shift of a blended
event to determine the degree of blending.  If the lensed star and the
blended objects are of different colors, then over the course of the
microlensing event, the color will shift from a weighted average all
of the contributing stars to approximately the same color of the
lensed star.  While this measurement does not give an unambiguous
measure of what the blending fraction is, it is nevertheless
indicative that a blend does exist.

Due to the narrow distribution of colors in the galactic bulge (Udalski et
al. 1993), it is difficult to detect significant color shifts in most events,
but development of early warning alert systems in both the OGLE (Udalski et
al. 1994b) and MACHO (Pratt et al. 1996) projects can permit hourly
observations of an event in progress with small photometric errors, making it
possible to detect a color shift.  Using a detailed model, Buchalter et
al. (1996) suggest that about $30 \%$ of bulge main sequence sources and about
$7 \%$ of bulge giant sources will show a shift using these uncertainties.  It
should be noted that the detailed models of color shifts concentrate almost
exclusively on blending by the lensing star, and suggests that crowding will
cause a color shift in only $\sim 10\%$ of events.

Han (1997b) proposes using the Hubble Space Telescope (HST) to do
followup analysis of a lensing event.  Since in the Galactic bulge
there will typically only be $\sim 4$ stars within a seeing disk at or
below 2 magnitudes above the detection limit, using the HST, one can
determine the positions and colors of the individual component stars.
By coupling this with the color information given by the chromatic
shift, one can select the star that is being lensed, and determine $f$
and $t_{0}$.

However, there are several difficulties.  First, the contribution of stars
fainter than 2 magnitudes above the detection limit is non-negligible.  The
model used indicates that there are $\sim 20$ stars in a typical PSF down to
$I=23$ magnitudes.  This is crowded even at HST resolution.  Since this limit
constitutes a fractional contribution of $0.01$ of the detection limit (at low
$f$, the probability of a lensed event being observed goes as $f$), but
increases the number of stars by an order of magnitude, these dim stars will
constitute $\gtrsim 10\%$ of the observed events.  Moreover, the HST time
involved is not inconsiderable ($\sim 1$ hr/event).

Finally, Alcock et al. (1996) discuss performing a statistical correction to
the timescales of observed events.  By performing Monte Carlo simulations, they
compute an estimate of the distribution of characteristic timescales, $t_{0}$,
under the assumption that $f=1$ throughout.  They can then multiply the
observed timescales by a correction factor in order to determine the optical
depth of the galactic halo to microlensing.  In principle this will give a good
overall estimate of the optical depth to microlensing, $\tau$, which is, after
all, the ultimate goal of these searches.  However, it does not give
information about individual microlensing events.  Furthermore, it requires a
detailed model of the stellar distribution in the Galaxy.

There has also been some discussion of a positional shift over the course of a
microlensing event.  Indeed, Alard et al. (1995; Alard 1996a) were able to
measure astrometric shift in DUO \# 2, confirming that it was blended.
However, since measurements were taken on a Schmidt plate, the shift was
measured by using a center of gravity, rather than a PSF fit, and hence, the
centroid position was only known to $\sim 0.2$ pixels at minimum light.

This technique has not been used more extensively, and what little attention
that it has received has dwelt on the idea that if anything, astrometric shifts
will be rare (Han 1997a).  In this paper, we propose that astrometric shifts
may be an extremely straightforward and robust method for searching for blends
in microlensing events, and potentially even for constraining the blending
parameter, $f$.  As this is a relatively new technique, we shall try to deal
with the problem in generality.  That is, rather than concentrate on a
particular survey or field, we shall use a simple (i.e. power law) luminosity
functions, with known degrees of crowding.  We think that this gives us a
handle on the frequency and importance of astrometric shifts in a way that may
be adapted to more realistic luminosity functions in a straightforward way.

The outline of the paper is as follows.  In \S 2, we discuss the reasoning
behind the expected astrometric shift, and show that this effect will be
significant in crowded fields.  In \S 4, we present a simulation of
microlensing events with a power law luminosity function within artificial CCD
frames.  In \S 3, we present the results of those simulations, and in \S 4, we
discuss the results, and suggest some future prospects.

\section{Motivation}

\subsection{Why do we expect an Astrometric Shift?}

\begin{figure}
\centerline{\psfig{figure=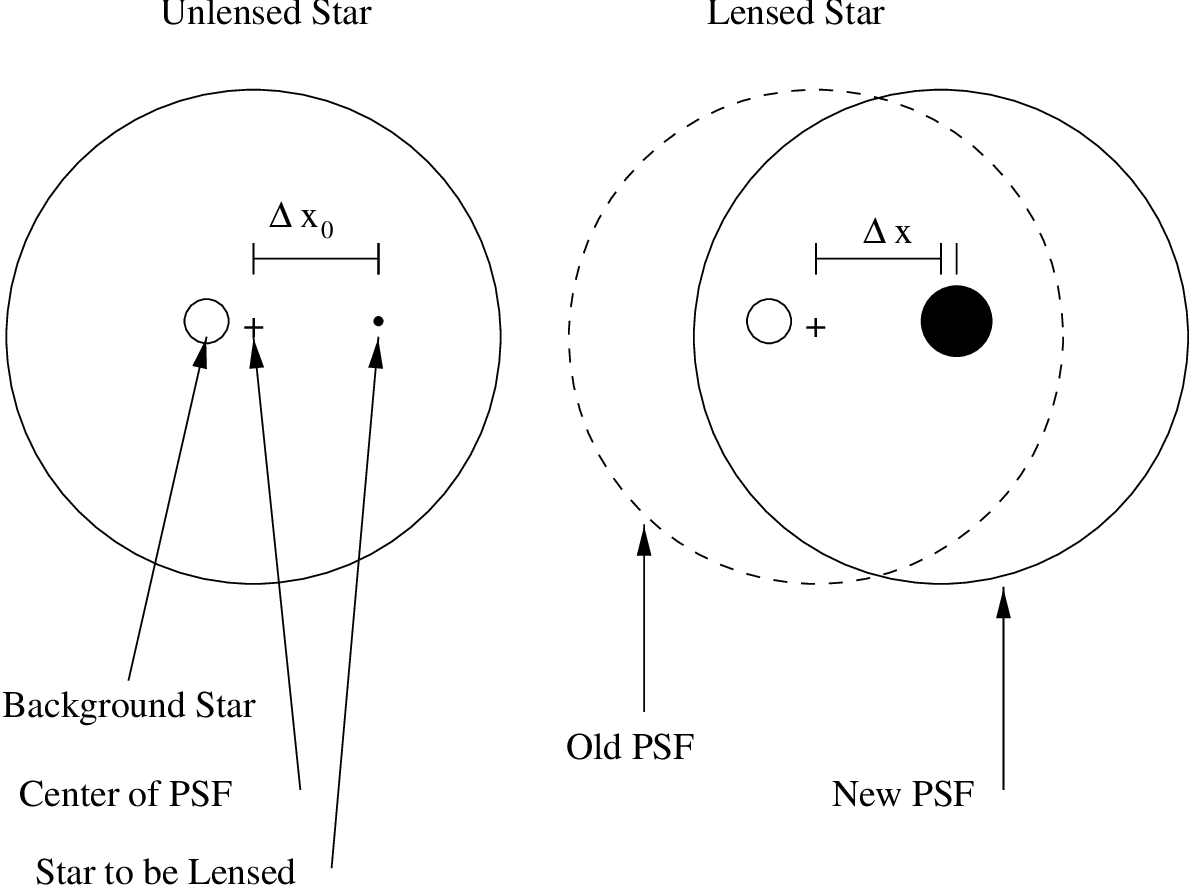,height=3.0in,angle=-90}}
\caption{A schematic description of the shift of the center
of light during a blended microlensing event.}
\label{fg:scheme1}
\end{figure}

Why do we expect to observe an astrometric shift in a blended microlensing
event?  Let us imagine that we only have a single, dark, lens which lenses one
of two blended stars.  For simplicity, we will assume that the PSF is a tophat
function, and hence, the center of the best fit PSF (and the estimated center
of light of the observed image) can be determined by a weighted average of the
two components.  If the lensed star has a fractional brightness, $f$, and is
some distance $\Delta x_{0}$ from the center of light, then the companion
contributes $(1-f)$, and is thus a distance $-\Delta x_{0}\ f/(1-f)$ from the
center of light.  If the lensed star increases in brightness by a factor, $A$,
the center of light will shift to:
\begin{equation}
\Delta x=
\Delta x_{0}\frac{f(A-1)}{f(A-1)+1}=\Delta x_{0}\frac{A_{obs}-1}{A_{obs}}\ ,
\label{eq:shiftline}
\end{equation}
where $A_{obs}$ is defined as above.  This is illustrated in
Figure~\ref{fg:scheme1}.  Though this relation assumes that there are only two
contributing stars, in fact, this should hold for any number of stars, since we
may put the stars into two groups: the lensed star, and all of the others.  If
we add the brightnesses and determine the centroid of the non-lensed stars,
then this combined object should essentially contribute to the observed
centroid like a single star.

One of the advantages of looking for an astrometric shift is that, unlike a
color shift, it is does not depend critically upon the distribution of stars.
Although the shape of the luminosity function and crowding may play a role in
determining how many events are blended, the presence of a shift essentially
only gives information about the relative geometry of the lensed source and the
observed image, and thus requires no particular stellar population model.  Thus
it is not a statistical correction, and does not depend on unobservables.

Moreover, it is inexpensive.  In current analysis of microlensing searches, a
template catalog provides constant central positions for each of the observed
objects, and the PSF for a given observation is determined from a set of
fiducial stars (eg Alcock et al. 1996).  While current methods may be ideal for
identifying microlensing events, they are not directly sensitive to astrometric
shifts.  Therefore, we propose using the template catalogs as well as the
frames containing the candidate events in order to do a PSF fit on the lensed
stellar image.  Since only a small fraction of the total frames will contain a
lensed event, this is far less computationally taxing than performing PSF
fitting throughout.  However, such data is not yet available for any of the
microlensing searches.  The OGLE database will soon be available in the form of
small subframes for the lensing candidates (Woz\'{n}iak \& Szyma\'{n}ski 1997)
and in a forthcoming paper, Goldberg \& Wo\'{z}niak (1997) will look for shifts
in the OGLE database.

Finally, even barring the possibility of finding a clear statistical relation
between $f$ and $\Delta r_{0}$, (where $\Delta r_{0}^{2}\equiv \Delta
x_{0}^{2}+\Delta y_{0}^{2}$), by performing a best fit to
equation~(\ref{eq:shiftline}), we can determine the position of the lensed star
with respect to the center of light.  In this way, high resolution,
ground-based followup observations can be used to observe the several stars
comprising the light within the PSF.  Since we can determine the position of
the lensed star within that PSF, it becomes a simple matter to determine its
fractional brightness compared to the observed PSF.

We suggest that it may be possible to detect a shift in a large percentage of
microlensing events.  Let us consider a case where there are exactly two stars
within the PSF, one much brighter than the other.  The PSF is very close to the
center of the PSF.  The other, however, may be anywhere within the PSF with
little effect on the centroid position.  If the dim object is lensed, we will
expect to see a shift in the centroid.  If the probability of finding the
center of the dim star is distributed randomly within $1\ R_{PSF}$, and the
centroid position can be measured to an accuracy of $\sigma_{r}$, then the
probability that there will be a measurable shift is
$1-(\sigma_{r}/R_{PSF})^{2}$, if $A_{obs}$ is large. For reasonable values of
$R_{PSF}=1$ pixel and $\sigma_{r}=0.1$ pixel, we expect that $\sim 99\%$ of the
events will have a measurable shift.

For an unblended (or very weakly blended star), however, we expect to see no
shift, since the centroid of the observed star is {\em defined} by the center
of the lensed star. 

\subsection{An Estimate: The Tophat Approximation}

Using a series of simplified assumptions, we can create a general order of
magnitude technique for determining how often we expect to see astrometric
shifts.  The purpose of this exercise is twofold: First, this approximation is
quick, and hence we can readily probe a family of luminosity functions.
Second, we can use this approximation as a ``sanity check'' on the more
rigorous simulations to follow.  We would like to reassure ourselves that this
effect is robust to our choice of parameters, and our choice of analysis
software.

For the ``Tophat Approximation'', we assume a tophat PSF of some fixed radius
(typically 3 ``pixels'', though the choice is arbitrary), and a power law
luminosity function, $n(L)\propto L^{-\beta}$, where $\beta$ will be a free
parameter.  In fact, we will consider this as our flux distribution, since the
lensed sources are approximately equidistant.  For clarity, this distribution
corresponds to a constant number per unit magnitude if $\beta=1$.  

We use a power law luminosity function throughout.  Though C. Alard (private
communication) has suggested that a power law provides a poor fit in the
galactic bulge (see eg Holtzman et al. 1993; Alard 1996b), it is not our
purpose to provide a field specific estimate, but rather to suggest that
astrometric shifts should be widely visible for a range of luminosity
functions.  Future work may wish to include more detailed models.

We model $N$ real stars per observed object (that is, blended within a given
PSF or radius $R_{psf}$), and we will set our detection threshold such that
there are $M^{-1}$ observed objects per unit PSF area.

For each star within a sufficiently bright PSF, we assign a value of $f$ given
by equation~(\ref{eq:fdef}).  We then convolve the distribution of $f$, with a
probability function.  That is, if a star with a given blending fraction, $f$,
were lensed, what is the probability that $A_{obs}>1.34$ ($u_{min}<1$)?  This
can be analytically determined and is not dependent upon our model, since from
simple geometrical arguments, the probability distribution of $u_{min}$ is
expected to be uniform:
\begin{equation}
P(A_{obs}>1.34| f)=\int_{1.34}^{\infty}p(A_{obs}|f)dA_{obs}=
\sqrt{\frac{2A_{min}(f)}{\sqrt{A_{min}^{2}(f)-1}}-2}\ ,
\end{equation}
where $A_{min}(f)\equiv 1+0.34/f$.

For each of the ``observed'' events, we can then determine a distribution of
offsets from the centroid.  Let us suppose that there is a group of stars with
a known center of light and a known brightness, and let us further suppose that
we then lay down an additional star, with a fractional brightness, $f$,
randomly with respect to the initial group.  It is simple to show that the
probability that the additional (lensed) star will be placed a distance, $r'$
from the initial centroid is simply $p(r')dr'=2r'/R_{psf}^{2}\ dr'$.  Thus, the
distribution of the distance of the lensed star from the center of light,
$\Delta r_{0}$, can be approximated as:
\begin{eqnarray}
p(\Delta r_{0})d(\Delta r_{0}) \simeq \frac{2(\Delta
r_{0})}{(1-f)^{2}R_{psf}^{2}} d(\Delta r_{0})\ &;&\Delta r_{0}<R_{psf}(1-f)\ .
\end{eqnarray}

This distribution is clearly correct in the limiting cases.  If $f\ll
1$, we expect that the star can be virtually anywhere in the PSF, and by virtue
of the larger area near the edges of the PSF, we will expect the lensed star to
have a high value of $\Delta r_{0}$.  If $f=1$, the star must be at the
centroid.

\begin{figure}
\centerline{\psfig{figure=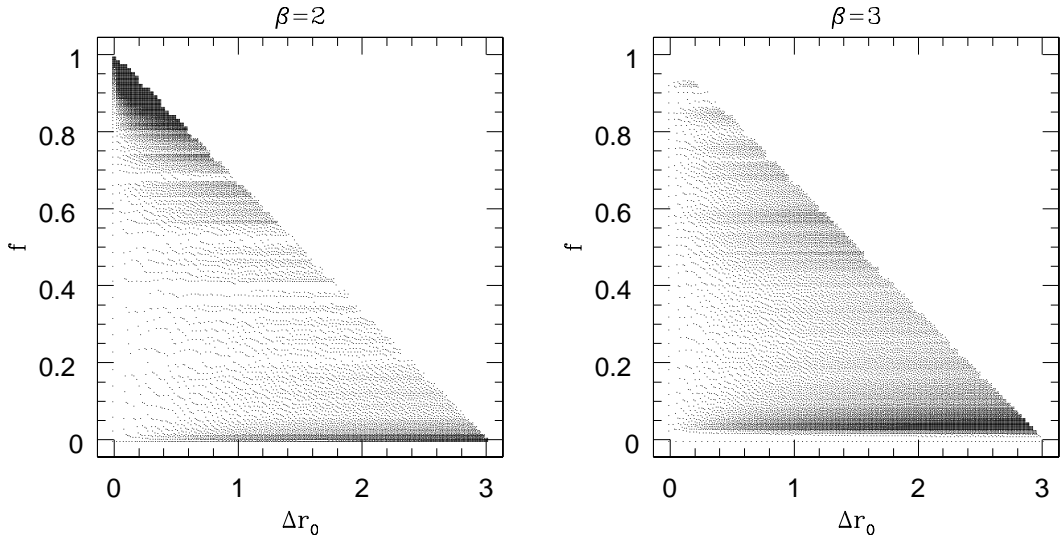,height=6in,angle=0}}
\caption{Covariant probability density of $f$ and $\Delta r_{0}$ for
the $\beta=2$ and $\beta=3$ luminosity functions.}
\label{fg:tophat2}
\end{figure}

Thus, we have a probability distribution of $f$ and a probability distribution
of $\Delta r_{0}(f)$.  We can relate these two, and in Figure~\ref{fg:tophat2},
we show a greyscale plot of the probability density of $p(\Delta r_{0},f)$, for
simple models with $N=10$ stars per PSF, $M=10$, and (a)$\beta=2$,
(b)$\beta=3$.  

\begin{figure}
\centerline{\psfig{figure=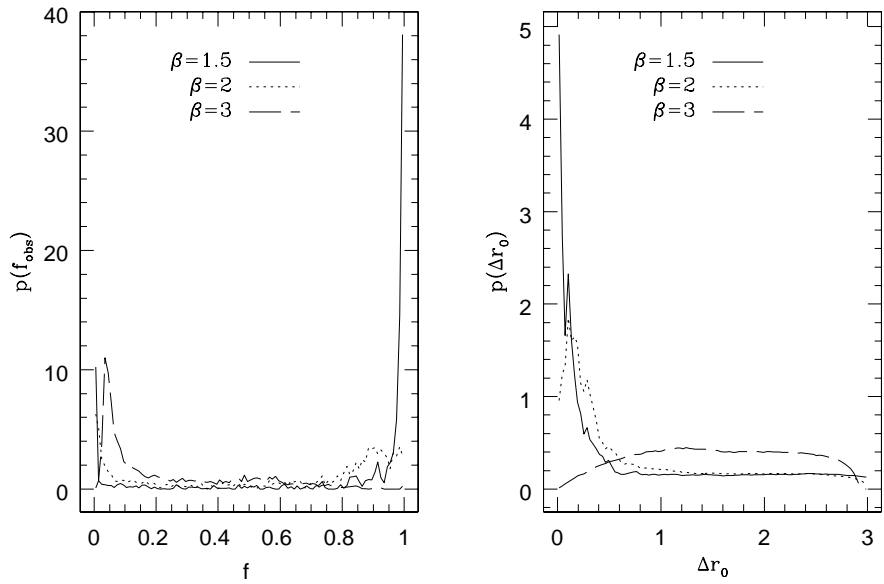,height=5in,angle=0}}
\caption{The expected distribution of $f$ and $\Delta r_{0}$ using the
``Tophat Approximation'' for a variety of luminosity functions.  The
luminosity functions are each power laws with $n(L)\propto L^{-\beta}$, where
$\beta$=$1.5,\ 2,$ and $3$, and $N=10$, $M=10$ in all three.}
\label{fg:tophat1}
\end{figure}

In Figure~\ref{fg:tophat1}, we show the expected distributions of $f$ and
$\Delta r_{0}$ for observed microlensing events with $\beta=1.5,\ 2$ and $3$.
For the $\beta=2$ and $\beta=3$ case, these are integrations of
Figure~\ref{fg:tophat2} over $\Delta r_{0}$ and $f$, respectively.

Note that for small values of $\beta$, the distribution is made up almost
entirely of unblended ($f=1$), and severely blended, ($f\ll 1$), events.  If
events were distributed in this, photometric analysis could differentiate
between the two (Wo\'{z}niak \& Paczy\'{n}ski 1997).  However, at higher
$\beta$, there are a great many events at intermediate blending fraction.  It
is these cases in particular that we hope to probe in our simulation.

Using this estimate, we find that many events are expected to show a
significant shift.  For $\beta =1.5, \ 2,$ and $3$ respectively, we find that
in very crowded fields, $0.38$, $0.43$, and $0.89$ of the events are expected
to have shifts of $\geq 0.2$ pixels ($\sim 0.1\times$ FWHM of the PSF).  These
typically represent $\sim 0.68-0.89$ of the events with $f<0.9$.

\section{Method: A Simulation of Blended Events}

Thus far, we have given only an approximation of the blending in the case where
we have perfect astrometry, no Poisson noise, an idealized PSF, and a complete
catalog below some limiting magnitude.  We now discuss a series of Monte
Carlo simulations in which we try to take more physical constraints into
account.

We first define a few terms, to which we refer throughout.  We use the term
``true'' catalog to refer to the actual positions and brightnesses of stars
which we use to generate a mock CCD image.  The ``observed'' catalogs are the
associated centers and brightnesses for observed objects computed by {\bf
DAOphot} package in {\em IRAF} (Stetson 1987).  The ``template'' catalog is
the collection of the most consistently identified objects over a series of
many observations.  Finally, a ``lensed'' catalog contains a collection of
observed objects if we alter the brightness of one or more true star.

We first create a true catalog with a power law luminosity function.
Each star is given a random position in coordinates of a $256\times
256$ grid with the limits of floating point accuracy (corresponding to
the dimensions of a CCD).

After creating a true catalog, we lay down a series of 15 observed images in
order to construct a template catalog.  First, the stars are laid down on a
mock CCD frame with a Gaussian PSF of $\sigma\simeq 1.2$ pixels.  Since there
are random shifts between different observations of a field, a random offset
of a fraction of a pixel is applied to all the stars before laying down a
template image.  A constant background of $\sim 2000$ counts per pixel
(compared with a typical background of $500-1000$ in the OGLE survey;
Wo\'{z}niak \& Szyma\'{n}ski 1997) and Poisson noise are added, and all pixels
above the saturation limit are truncated at that limit.  We assume no read
noise or bad pixels.  The former should not be a major consideration since
crowded fields are confusion, rather than photon limited, and the latter should
effect all events democratically.

Each of the observed images are then analyzed using {\bf DAOphot}, and in
particular, brightnesses and centers of each star in each of the observed
catalogs are computed using a PSF fitting routine.  We assume that since there
are a large number of stars in the frame, the width of the PSF is well known.
We then cross compare each of the observed catalogs, and retain only those
stars which appear in all 15 of the observed catalogs.  The template catalog is
the set of mean brightnesses, central positions and uncertainties, of these
consistently observed stars.

We then proceed to compute a series of lensed catalogs.  For a series of
amplifications, we brighten each of the stars that appear within the PSF of a
given observed object sequentially, and use {\bf DAOphot} to create an observed
catalog.  For reasons of computational speed, we adjust the brightness of one
true star in each of the template objects simultaneously.  We then compare the
lensed catalog to the template catalog to see whether each of the lensed
objects is observed, and whether changing $A$ actually affects $A_{obs}$.
Since we have only the physical proximity of the true and observed star to
suggest that they are related, the only way to test whether the two are
observationally associated is to see whether varying the former also changes
the latter.

For those true stars associated with an observed star, we create a series of
mock microlensing events.  We select a random $u_{min}$ between zero and one
for each star.  If $A_{max}^{obs}\geq 1.34$ (corresponding to $u_{min}=1$ for
an unblended event), we include this event in our statistics.

Thus, for each event, we have a set of values of $A_{obs}$, $\Delta
x$, $\Delta y$, and $A$, where the former 3 are observables, and the
latter is only known in a computational environment.  Typically, we
have $\sim 5-10$ measurements per event, about $\sim 0.2-0.5$ of the
number in current microlensing searches.

We generate a mock light curves, by plotting $A_{obs}$, versus $t$ (which can
be computed from the known ``true'' magnification of the star).  We also use
our catalogs to compute $\Delta x_{0}$ and $\Delta y_{0}$, where $\Delta x_{0}$
is the best fit to equation~(\ref{eq:shiftline}) with a similar equation for
$\Delta y_{0}$.  By definition, $\Delta x_{0}$ and $\Delta y_{0}$ are
equal to the position of the true lensed star with respect to centroid of the
observed star at baseline.  Finally, we determine the fraction of light in our
lensed object, $f$, by doing a best fit to equation~(\ref{eq:magline}).

\begin{figure}
\centerline{\psfig{figure=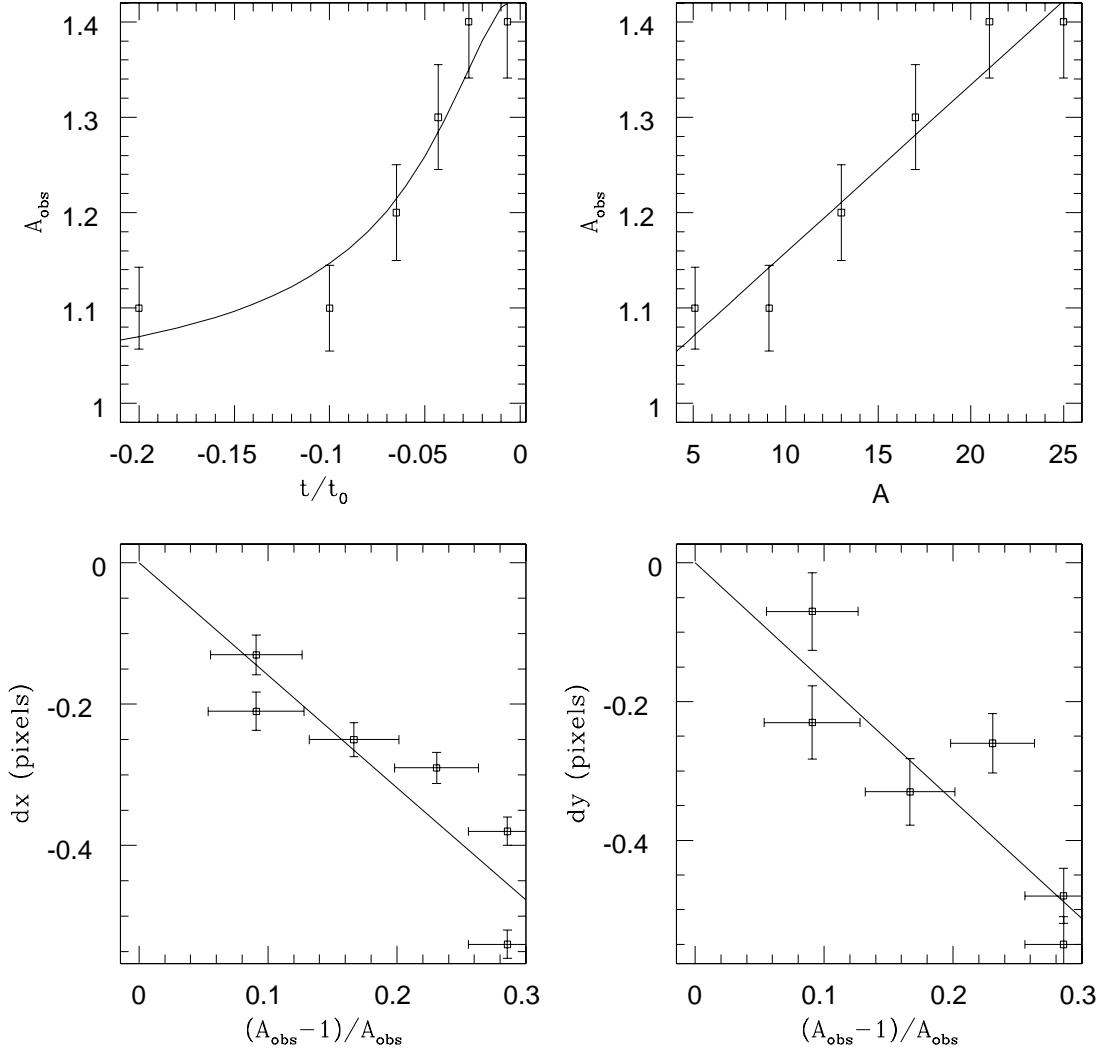,height=6in,angle=0}}
\caption{A mock microlensing curve, with $f=0.018$ and
$u_{min}=0.04$. (a) The photometric light curve with associated errors.
The line is the curve associated with the microlensing parameters.
(b)A plot of Magnification, $A_{obs}$, versus Amplification, $A$.  The
line is a best fit linear relation given by equation~(\ref{eq:magline}).
(c)The astrometric shift of the measured PSF centroid in the $x$ coordinate.
The line is given by equation~(\ref{eq:shiftline}), and corresponds to an
offset of $\Delta x_{0}=\simeq -1.6$ pixels.  (d) Similar to panel c, but in the $y$ 
coordinate.  The measured offset is $\Delta y_{0}=-1.7$ pixels.}
\label{fg:curve1}
\end{figure}

In Figure~\ref{fg:curve1}, we show a typical parameter fit to a simulated
microlensing event with $f_{obs}=0.018$, $\Delta r_{0}=2.3$ pixels, and
$u_{min}=0.04$.  This corresponds to an $A_{obs}\simeq 1.4$.  Note that in
accordance with our initial assumptions, a strong blend is also accompanied by
a large shift.

\section{Results}

In this section, we present the results of our simulations.  We ran two
simulations, each with 20000 stars.  We used a power law luminosity function
for both, one with $\beta=2$, and the other with $\beta=3$.  The former
contained $\sim 650$ template objects, and the latter contained $\sim 400$
objects.  In both cases, about twice as many stars were typically observed, but
many were rejected from the template catalog due to the fact that they were not
observed in all of the catalogs.  Since both fields were limited by crowding, a
typical observed star was associated with $\sim 15$ blended true stars, down to
about $1\%$ of the observed threshold.  This is consistent with the model used
by Han (1997b).  We verified that the observed template objects had a high
correspondence to the brightest stars in the true catalog.

In order to compare the simulations to actual observational programs, we will
briefly consider the crowding of the template catalog.  The template catalogs
contain $\sim 1$ star per $100$ pixels.  Though typical observed catalogs were
about twice as crowded.  By comparison, the OGLE survey (Udalski et al. 1992)
has an average of $\sim 1/25$ pixel, while the MACHO LMC survey (Alcock et
al. 1996a) has an average of $\sim 1$ monitored LMC star per $80$ pixels.
Future simulations may wish to model specific fields.

After creating a set of lensed catalogs, and randomly selecting impact
parameters, $u_{min}$ for each of the consistently observed, lensed objects,
we found $\simeq 500$ microlensing events in the $\beta=2$ simulation, and
$\simeq 600$ events in the $\beta=3$ simulation.

\begin{figure}
\centerline{\psfig{figure=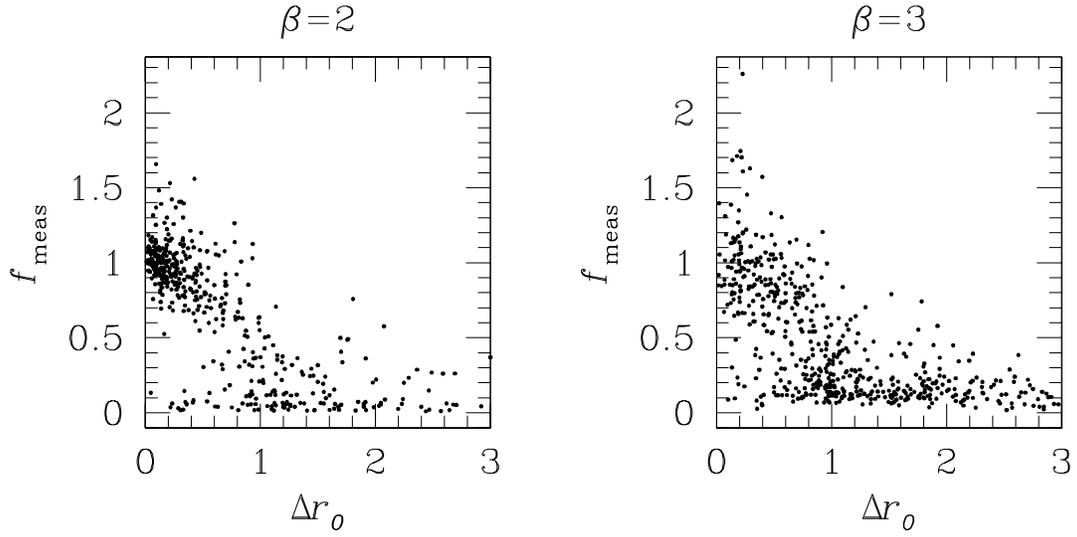,height=6in,angle=0}}
\caption{A scatter plot showing the relation between blending fraction,
$f$, and measured offset, $\Delta r_{0}$ for the (a)$\beta=2$, and (b)
$\beta=3$ case.}
\label{fg:scatter1}
\end{figure}

In Figure~\ref{fg:scatter1}, we show the results of these simulations.  Panel
(a) shows the distribution of $f$ (as fit to equation~\ref{eq:magline}) against
$\Delta r_{0}$ (as fit to equation~\ref{eq:shiftline}) for the $\beta=2$
simulation.  Panel (b) shows the distribution for the $\beta=3$ case.  Note
that the simplistic tophat approximation reproduces the gross features in the
actual distribution of events (Figure~\ref{fg:tophat2}), especially for $\beta
= 2$.

Some of the events are ``measured'' as having $f>1$.  There are two reasons for
this: First, the fits are based on measured data, and hence, any uncertainties
in the data produce uncertainties in $f$.  Within errors, many $f=1$ events can
certainly be measured at $f>1$.  Second, there is a significant background
effect in crowded fields.  Some of the light of the lensed star may be
considered part of the background in the unlensed case, and hence, the increase
in brightness (which can be determined differentially), can appear relatively
more substantial than, in fact, it is.  We shall consider these $f>1$ events to
be unblended for the rest of the analysis.

Note also that the $\beta=2$ model has many unblended events with essentially
no shift, while the $\beta=3$ model seems to have a larger fraction of events
which have are severely blended.  In the former case, we think of most observed
stars being identified as a primary, and the many other contributing stars can
only produce visible events by being in close proximity.  On the other hand, in
the $\beta=3$ case, the more democratic distribution of values of $f$ suggest
that, due to the greater number of dim stars (from the steeper luminosity
function), many observed ``stars'' contain no single true star which would be
of sufficient brightness to be visible on its own.

In addition to the distribution of shifts and blending fractions, we are also
interested in the relation between the two.  Table 1 lists the number of events
which were observed with significant shifts, compared with whether or not they
were significantly blended.  Note that in both simulations, $\sim 75\%$ of
significantly blended events had a large shift $(\Delta r_{0}>0.5 \ {\rm
pixels})$, while only $\sim 3\%$ of unblended events had a large shift.

\begin{figure}
\centerline{\psfig{figure=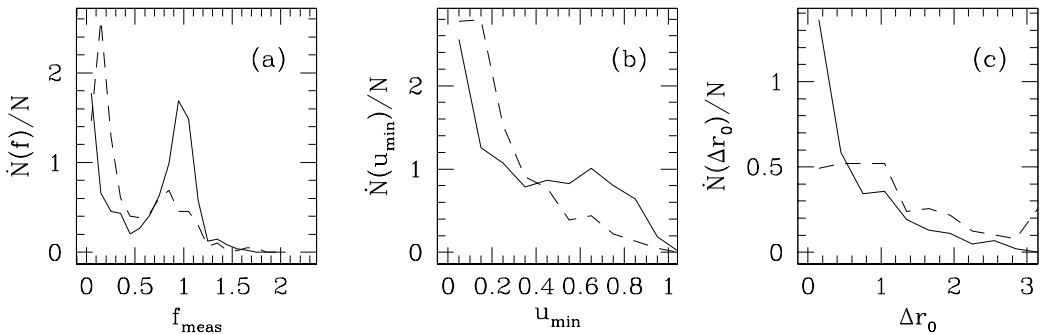,height=6in,angle=0}}
\caption{ The results of the $\beta=2$ (solid) and $\beta=3$ (dashed)
simulations. (a) The distribution of microlensing events as a function
of $f$, binned into $\Delta f=0.2$ bins.  (b) The distribution of
microlensing events as a function of $u_{min}$, binned into $\Delta
u_{min}=0.1$ bins.  (c) The distribution of events as a function of
astrometric shift, with data binned into $\Delta r_{0}=0.3$ pixel
bins.}
\label{fg:lens1}
\end{figure}

We consider the distribution of $f$ and $\Delta r_{0}$ and $u_{min}$, for each
of the models, in order to see how often we expect to see shifts, or to
estimate how often events will be significantly blended.
Figure~\ref{fg:lens1}a shows the distribution of values of $f$.
Figure~\ref{fg:lens1}b shows the distribution of $u_{min}$.  It is strongly
skewed to low values of $u_{min}$ due to events with small $f$, which will not
be detected if the true value of $u_{min}$ is close to unity.
Figure~\ref{fg:lens1}c shows the distribution of shifts.

Figure~\ref{fg:lens1}a is of considerable interest.  In both models, there seem
to be two populations of events.  The first population consists of essentially
unblended ($f\simeq 1$) events, while the second group are highly blended
($f\ll 1$), resulting in a bimodal distribution of blending fractions (with a
typical FWHM of $\Delta f\sim 0.5$).  In a sparse field the population of
highly blended events would not be seen at all, since their relative importance
is related directly to the number of stars below the observational threshold
within the PSF of an observed image.  This two component model may also
describe a more realistic luminosity function.

In an actual observational program, we would only be able to measure
Figure~\ref{fg:lens1}c.  However, even from this, we are able to make some
estimates about the blending fraction.  For example, we note that that about
twice as many events have a small astrometric shift ($\Delta r_{0}<0.5$
pixels), in the $\beta=2$ model as the $\beta=3$ model.  To first order, one
would think (correctly) that there will be about twice as many unblended events
in the $\beta=2$ model as the $\beta=3$ model.  Indeed, since the distribution
of shifts (and with it, the distribution of $f$) is strongly dependent upon the
luminosity function of stars below the detection threshold, this distribution
may be a powerful probe of the stars in that regime.  For a large survey like
MACHO, with $>100$ observed events, this distribution may actually be observed.

For a given scenario (Luminosity Function+Crowding+Detection Method)
we can compute (numerically) the probability of observing an event
with blending fraction, $f$, given some observed astrometric shift,
$p(f|\Delta r_{0})$, and likewise determine the probability
distribution of astrometric shifts given some ``true'' blending
fraction: $p(\Delta r_{0}|f_{0})$.  These probabilities are
appropriately normalized vertical and horizontal traces through
Figure~\ref{fg:scatter1}.  Both dispersions are quite large.  However,
we may ask: if an event has a true blending fraction, $f_{0}$, what is
the probability that we will estimate the blending fraction, $f$?
\begin{equation}
P(f|f_{0})df=\int_{\Delta r_{0}=0}^{\infty}P(\Delta r_{0}|f_{0})P(f|\Delta r_{0})d(\Delta r_{0}) df\ .
\end{equation}
Thus, we can calculate the expectation value, $f_{meas}$, and the associated
dispersion, given some actual blending fraction, $f_{0}$.  

\begin{figure}
\centerline{\psfig{figure=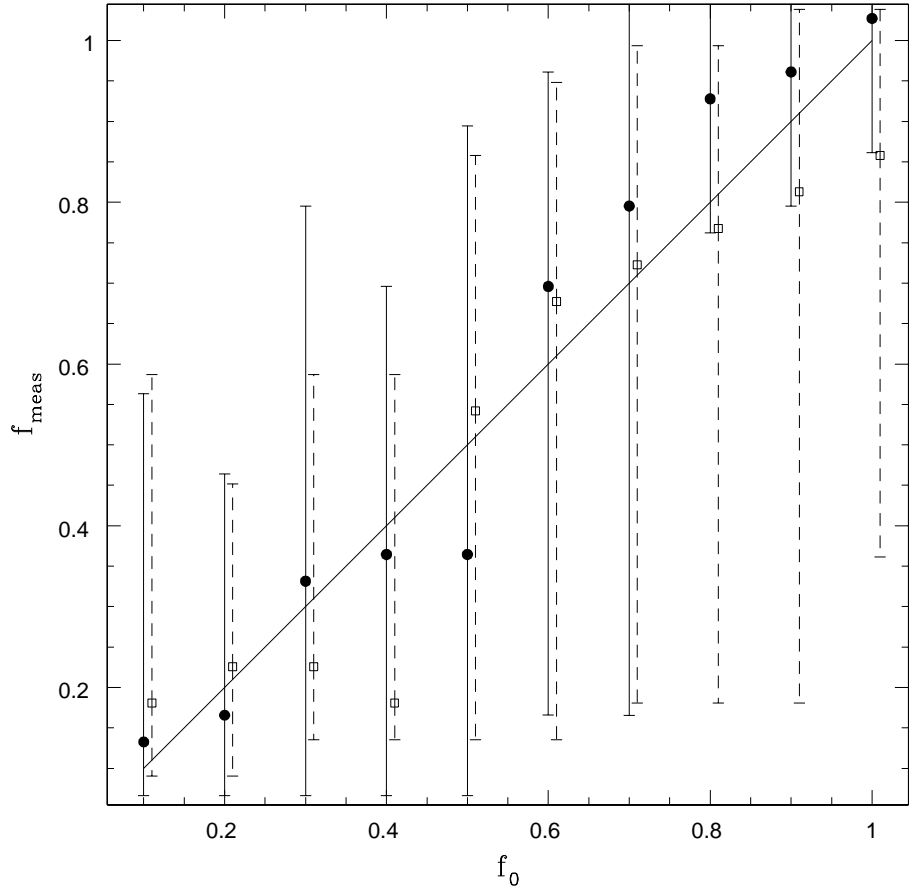,height=5in}}
\caption{The distribution of actual and statistically measured values of
$f$.  The solid line shows an unbiased estimate of $f$.  The solid circles
and associated errors show the median and $1-\sigma$ dispersion of the
estimated $f$'s for the $\beta=2$ simulation, while the open squares show
the probability distribution for the $\beta=3$ case}
\label{fg:fmeasure}
\end{figure}

In Figure~\ref{fg:fmeasure}, we present this result.  The extreme
cases of an unblended or a highly blended ($f\lesssim 0.1$) event can
be statistically distinguished from one another.  However, moderate
blends ($f\simeq 0.5$) are generally consistent with either extreme.

\section{Discussion}

\subsection{Comparison with Previous Deblending Estimates}

Does this result constrain blending significantly?  Wo\'{z}niak \&
Paczy\'{n}ski (1997) suggested that, in general, photometric information was
insufficient to uniquely determine a value of $f$.  For $f\gtrsim 0.08$, a
significant fraction of events could not be distinguished from the unblended
case at the $68\%$ level.  By comparison, from Figure~\ref{fg:fmeasure}, we see
that by using astrometric shifts, an event can be distinguished at the $68\%$
confidence level if $f\leq 0.6$ for the $\beta=2$ model, and $f\leq 0.5$ for
the $\beta=3$ model.

Regrettably, modest values of $f$ are difficult to uniquely identify using
either astrometric shifts and light curve fitting, but in a slightly different
way.  Though essentially {\em all} significant shifts correspond to a
significant blend, we cannot say that all significant blends will have a
significant shift.

This method may also do quite well in comparison with the color shift method.
First, searches for astrometric shift require no {\em a priori} model of color
distribution.  Moreover, a noticeable astrometric shift was predicted for
$\gtrsim 0.6$ of the events in the $\beta=2$ model and $\gtrsim 0.8$ in the
$\beta=3$ model.  In both, those with small shifts $(\leq 0.5 \ {\rm pixels})$
were virtually all $(\geq 90\%)$ unblended, and most of those with large shifts
$(> 0.5 \ {\rm pixel})$ were severely blended.  This can be contrasted with the
$30\%$ color shift detection rate suggested by Buchalter et al. (1996) for
blending by the lens, and the $10\%$ shift rate for background blends.  They
assumed $\sim 0.1$ bright ($V<19$) stars per arcsecond in the bulge, and
assuming $1$ arcsecond seeing our distribution gives $\sim 0.06$ visible stars
per arcsecond, suggesting that our degree of crowding is not unreasonable for
current observations.

\subsection{Other Complications and Considerations}

We have used a fairly simple model of astrometric shift throughout.  There may
be considerable concern that we have used an overly simple luminosity function,
or an unphysical degree of crowding.  However, it was not our aim to deal with
a specific survey or a specific field, and thus we have tried to treat the
problem with generality.  However in addition to these types of concerns, there
are some additional complicating issues which we may wish to keep in mind.

The first of these is that we are not considering fixed positions for PSFs, and
hence, the weight given to various background stars will vary over the course
of a microlensing event.  As a result of this, the value of $f$ will vary, and
one might expect that equations~(\ref{eq:magline}) and~(\ref{eq:shiftline})
will no longer be strictly correct.  In our simulations, this has not posed a
problem.  A constant value of $f$ throughout satisfies both relations
adequately.  It is less clear that a moving centroid will provide identical
measured parameters as a constant centroid.  By following the center of light,
it may be that the amplification may be systematically higher than with a
constant centroid owing to an increasing contribution of the lensed star.

Next, this method is not directly sensitive to blending by a binary companion
since the angular separation between the two would be miniscule.  Given that
about $\sim 0.8$ of all bright primary stars are expected to have a companion
(Abt et al. 1990), this is not a small effect.  We can consider a worst case
scenario.  Imagine that all stars have a companion of equal brightness, but are
assumed to be single stars.  We will systematically overestimate $f$ by a
factor of $2$.  However, we will also underestimate the effective optical depth
for a given Galactic model, since each monitored star will represent two
chances to be lensed.  In a more benign case in which one star is brighter than
its companion, we will expect that there will be a bias toward observing the
brighter star being lensed.  Moreover, if it is significantly brighter than its
companion, then the effect of blending (by the binary) will be small.

Finally, we have other observed parameters which we have not used directly.  In
attempting to deblend an event, we can also specify the observed brightness in
our probability function.  Consider a very bright object in a region with a
steep luminosity function, for example.  We may expect that {\em a priori} the
star will be almost unblended, and thus, for a microlensing event which has a
baseline level many $\sigma$ above the mean, we are virtually guaranteed that
it is unblended.  Looking at the brightest $10\%$ of events in each of our
simulations, we found that in the $\beta=3$ (brightest $48$ events) only $4$
events had a value of $f<0.8$, and of those, only $3$ had a shift of greater
than $0.5$ pixels.  For a flatter distribution like the $\beta=2$ simulation
(brightest $56$ events), the distribution of brightest events more strongly
resembled the distribution of events as a whole, with $\sim 30$ events with
$f<0.8$ and shift $>0.5$ pixels.

\subsection{Future Prospects}

We propose a twofold investigation into astrometric shifts of blended
microlensing events, numerical and observational.

First, there is a need for more complex and survey specific simulations.  We
have used a pre-packaged (though commonly used) routine in order to determine
centroid positions.  More robust techniques exist.  In particular {\bf DoPHOT}
(Schechter et al. 1995) has been extremely successful in determining positions
in crowded fields.  Likewise, a more accurate and field specific luminosity
function, variable seeing, realistic distribution of observations, and so
forth, may make it possible to calibrate a simulation with a particular
observational program.

In conjunction with theoretical estimates, it is also hoped that
researchers will begin to look for astrometric shifts within current
observational programs.  This can be done with existing data after the
fact, and can be done extremely quickly if only a small subframe
around the lensed object is considered.  The OGLE data is being
prepared in this form (Wo\'{z}niak \& Szyma\'{n}ski 1997) and Goldberg
\& Wo\'{z}niak (1997) have detected a shift of $0.7$ pixels for OGLE
\# 5, an event already considered to be blended.

More importantly, perhaps, is the fact that astrometric information gives us an
unambiguous (and model independent!) determination of the centroid position of
the actual lensed star compared to the observed centroid of the PSF.  By
performing high-resolution ground based or HST followup observations, the
actual source star could be picked out of a very crowded field, and thus used
to directly deblend the event.

\acknowledgements{ The author would like to gratefully acknowledge the advice
and many valuable discussions with Bohdan Paczy\'{n}ski, as well as helpful
comments from Przemyslaw Wo\'{z}niak, Christophe Alard, Michael Strauss, and Michael Richmond.
This work was supported by NSF grant AST-9530478 and an NSF Graduate Research
Fellowship.}

\newpage

\newpage

\begin{table}[h]
\begin{tabular}{|c||c|c||c|c|}
\hline
\multicolumn{1}{|c}{} &
\multicolumn{2}{|c|}{$\beta=2$ (total=485)} &
\multicolumn{2}{|c|}{$\beta=3$ (total=495)}\\ \hline\hline
\multicolumn{1}{|c|}{}& $f>0.9$ & $f\leq 0.9$ & $f>0.9$ & $f\leq 0.9$ \\ \hline
$\Delta r_{0}>0.5$ & 14 & 206 & 20 & 427 \\ \hline
$\Delta r_{0}\leq 0.5$ & 188 & 77 & 71 & 77 \\ \hline
\end{tabular}
\caption{A summary of the distribution of events in the $\beta=2$ and
$\beta=3$ microlensing simulations.  The events are roughly divided
into ``unblended'' $(f>0.9)$, ``significantly blended'' $(f\leq 0.9)$,
``unshifted'' $(\Delta r_{0}\leq 0.5 \ {\rm pixels})$, and ``shifted''
$(\Delta r_{0}> 0.5 \ {\rm pixels})$.}
\label{tab:summary}
\end{table}

\end{document}